# La dignificación de la pepena: un análisis del reciclaje de residuos sólidos urbanos en la ciudad de Chihuahua

## The dignification of scrap-picking: an analysis of urban solid waste recycling in the city of Chihuahua

Terrazas-Chavira Gabriela Alejandra, Espino-Enríquez Lauro Manuel, Frescas-Villalobos Raúl Hiram, Manjarrez-Domínguez Carlos Baudel, Hoffmann-Esteves Hazel Eugenia

Universidad Autónoma de Chihuahua.

### NOTA SOBRE LOS AUTORES
Gabriela Alejandra Terrazas-Chavira: gaaltech06@gmail.com
https://orcid.org/0009-0008-6977-3745
Lauro Manuel Espino Enríquez: lespino@uach.mx
https://orcid.org/0000-0003-4122-5049
Raúl Hiram Frescas Villalobos: rhfrescas@uach.mx
https://orcid.org/0009-0000-4108-6427
Carlos Baudel Manjarrez Domínguez: mdmanjarrez@uach.mx
https://orcid.org/0000-0001-9536-4007
Hazel Eugenia Hoffmann Esteves: hhoffmann@uach.mx
https://orcid.org/0000-0002-8744-9606

## RESUMEN

La actividad de la pepena se encuentra dentro del sector informal, caracterizada por la precariedad e invisibilidad social, por el contrario, es un elemento clave en la cadena productiva del reciclaje, brindando un ingreso económico para muchas personas. En la ciudad de Chihuahua esta actividad se realiza a una escala importante, por lo que el presente estudio analiza y dignifica el reciclaje de los residuos sólidos urbanos desde la perspectiva de los pepenadores, explicando también su relación con otros actores sociales en el sitio de disposición final de la ciudad. Se parte de la premisa de que la pepena de basura es, en la época actual de crisis ambiental global una forma efectiva de cuidado de recursos y disminución de impactos ambientales. A nivel metodológico, se trabajó desde la investigación acción participativa, como estrategia de sensibilización en el grupo de pepenadores de la importancia de su labor y, una forma de presentar su organización comunitaria como ejemplo a otros grupos urbanos de pepena de basura en otras latitudes. Finalmente encontramos problemáticas en estos grupos como lo son: la informalidad, la falta de respaldo jurídico, la exposición a la salud de forma particular, mientras que de forma general, la falta de políticas públicas para reconocer esta actividad como valiosa para afrontar la crisis ambiental global.

**Palabras claves:** Pepenador, ecología urbana, crisis ambiental.

## ABSTRACT

The pepena activity is part of the informal sector, characterized by precariousness and social invisibility; however, it is a key element in the recycling production chain, providing an economic income for many people. In the city of Chihuahua, this activity is carried out on a significant scale. Therefore, this study analyzes and dignifies the recycling of urban solid waste from the perspective of scavengers, also explaining their relationship with other social actors at the city's final disposal site. The study starts from the premise that waste picking is, in the current era of global environmental crisis, an effective way to conserve resources and reduce environmental impacts. Methodologically, participatory action research was used as a strategy to raise awareness among scavengers about the importance of their work and to present their community organization as an example to other urban waste picking groups in different locations. Finally, we found issues in these groups such as informality, lack of legal support, specific health risks, and, more broadly, the absence of public policies to recognize this activity as valuable in addressing the global environmental crisis.

## INTRODUCCIÓN

Un número creciente de personas están habitando las ciudades cada vez más grandes. Estas comunidades de alta densidad poblacional plantean un reto especial en cuanto a la generación de residuos sólidos (Gary, 1999), llegando actualmente a representar uno de los retos más sobresalientes para la sociedad en su relación con el ambiente. Se estima que en el planeta se generan alrededor de 1600 millones de toneladas anuales de residuos sólidos (Angulo et al., 2010), con repercusiones ambientales tales como emisiones de metano y dióxido de carbono (Qdais et al., 2010), olores ofensivos en los rellenos sanitarios, contaminación atmosférica y de recursos hídricos.

El pepenador es fundamental en el proceso de recuperación de materiales para reciclaje,

rescatando una cantidad considerable de residuos de la disposición final. El reciclaje permite valorizar el potencial comercial, energético y de transformación del residuo para transformarlo en recurso e impactar en el ahorro de materias primas que se obtienen de los entornos naturales). Un estimado de 24 millones de personas alrededor del mundo participan en las actividades de reciclaje: recolección, recuperación, separación, clasificación, limpia, empaque, compactación y procesado de los residuos en nuevos productos (WIEGO, 2014)

Sin la aportación del pepenador la brecha en porcentajes de reciclado entre los países del primer y tercer mundo sería aún más grande. Esta actividad conserva los recursos y realiza un trabajo socialmente útil, económicamente productivo y ambientalmente benéfico (Lozano et al., 2009), que al mismo tiempo, suele encontrarse fuertemente estigmatizado, relacionando su actividad con la marginalidad.

Finalmente, el trabajo pretende dar visibilidad y dignificación a una actividad que contribuye a la reducción de impactos ambientales por la emisión de residuos sólidos urbanos. Para ello, se utiliza como caso de estudio de la Ciudad de Chihuahua en el sitio de disposición final de basura (relleno sanitario), analizando la estructura social, relación de actores, importancia y sensibilización para la dignificación de la pepena de forma inmanente.

## MATERIALES Y MÉTODOS

Área de estudio. Según datos del INEGI (2020), la población total de la ciudad de Chihuahua es de 937,674 personas, el crecimiento en su población refleja edificaciones de más conjuntos habitacionales lo cual repercute en la generación de residuos sólidos urbanos debido al aumento de colonias, establecimientos comerciales y de servicios, es por ello que el manejo de los residuos se realiza en el sitio de disposición final de la ciudad en el kilómetro 7.5 carretera a Aldama, el cual permite que más de 350 trabajadores realicen la actividad de la pepena, integrándose al porcentaje del sector informal.

Metodología de Intervención: Investigación Acción Participativa (IAP). La Investigación Acción Participativa (IAP), considerada como una metodología aplicada en contextos sociales, discurre de manera convergente en un plan de acción a partir de las características de la cultura y la organización social. El artículo de Balcazar, F. E., (2003), La IAP provee un contexto concreto para involucrar a los miembros de una comunidad o grupo en el proceso de investigación en una forma no tradicional como agentes de cambio y no como objetos de estudio. Fue seleccionada por ser una metodología mixta de alcance transversal, por sus bondades que brindan rigidez a los datos duros y por la profundidad en la intervención cualitativa.

En materia al presente estudio, la IAP es aplicada con intervención específica en los pepenadores de los residuos sólidos urbanos generados en la ciudad de Chihuahua, con el objetivo de realizar un estudio diagnóstico de su situación social, dignificar su actividad en la comprensión interna del grupo y obtener de forma transversal información que justifique la importancia de la actividad. Para el logro del objetivo anterior se buscó el paso de la relación sujeto-objeto a la relación sujeto-sujeto como estrategia de apropiación consciente de su solución y cambio. Las etapas de la IAP son estructuradas acorde a las necesidades del espacio que se interviene, se considera para este trabajo secuencia metodológica de tres etapas generales, siendo sus momentos principales los mostrados en la figura 1:

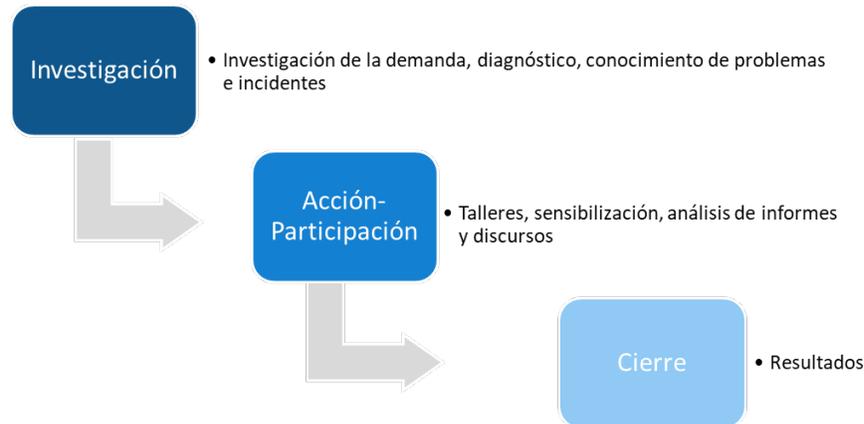

Figura 1. Etapas de la metodología de intervención- Investigación Acción Participativa.

Etapa 1 Investigación. - Se realizó un mapeo de actores con la finalidad de identificar a los participantes de mayor relevancia para el presente estudio, mismos, que como resultado podían haber sido personas de la comunidad, funcionarios de instituciones públicas, representantes de empresas, asociaciones privadas, organizaciones sociales y usuarios. Seguido a la identificación del mapa de actores se contactó a los mismos, llevando a cabo acercamientos con entrevistas semi estructuradas y a profundidad, logrando así la aceptación y el apoyo al estudio.

Una vez teniendo la aceptación, se desarrolló un instrumento de investigación para recolectar información sobre la situación de los actores involucrados en la recolección de residuos sólidos urbanos. La encuesta fue diseñada para la generación de información diagnostica (radiografía) en la relación sujeto-objeto; constando de 24 preguntas abiertas y cerradas.

Para la aplicación del instrumento se calculó una muestra con un margen de error del 5% y se consideró un universo de 380 personas, resultando una muestra de 191.28 personas a encuestar.

Etapa 2 Acción-Participación. - En esta etapa fueron aplicados dos instrumentos: El primero, que consiste en la encuesta descrita en la etapa uno. El segundo consistió en talleres participativos con actores principales con el objetivo de llevar a cabo la socialización y pasar a la relación sujeto-sujeto para la intervención de la solución al problema. Durante el primer instrumento y siguiendo el resultado de la muestra anteriormente presentada, se logró trabajar con 201 pepenadores buscando prevenir errores en el proceso de entrevistas y así converger al nivel de confianza propuesto.

La información recolectada se capturó en bases de datos por temáticas del instrumento el cual, integró las situaciones sociales, salud, educación y cultural que tiene cada actor, lo cual permitió generar posibles recomendaciones para aplicación en el estudio.

Etapa 3 Cierre. - Se analizaron los resultados obtenidos durante el desarrollo de las dos etapas anteriores y con ellos se identificaron posibles rutas de acción para favorecer la práctica de la pepena en otros espacios.

## RESULTADOS

Con la identificación realizada en la etapa 1 se obtuvo el siguiente mapa de actores (figura 2), donde se puede analizar el papel de cada actor, así mismo, se observan tres niveles y su área de influencia, lo cual visualiza las relaciones directas e indirectas entre ellos. Así mismo, demuestra la jerarquización que existe en los pepenadores dentro de su organización.

El área de influencia es distinta en los tres niveles, en el área de color azul se representa el espacio físico que comparten en el sitio de disposición final. En el área de color verde se encuentran los recolectores primarios. Finalmente, en el área de color amarillo es ocupado por los recolectores viales. El mapa nos demuestra que ningún actor contraviene las actividades de otro, no compiten por los residuos y de manera informal tienen conocimiento de esta estructura orgánica, brindándole un alto respeto simbólico llamándola entre ellos "mesa directiva".

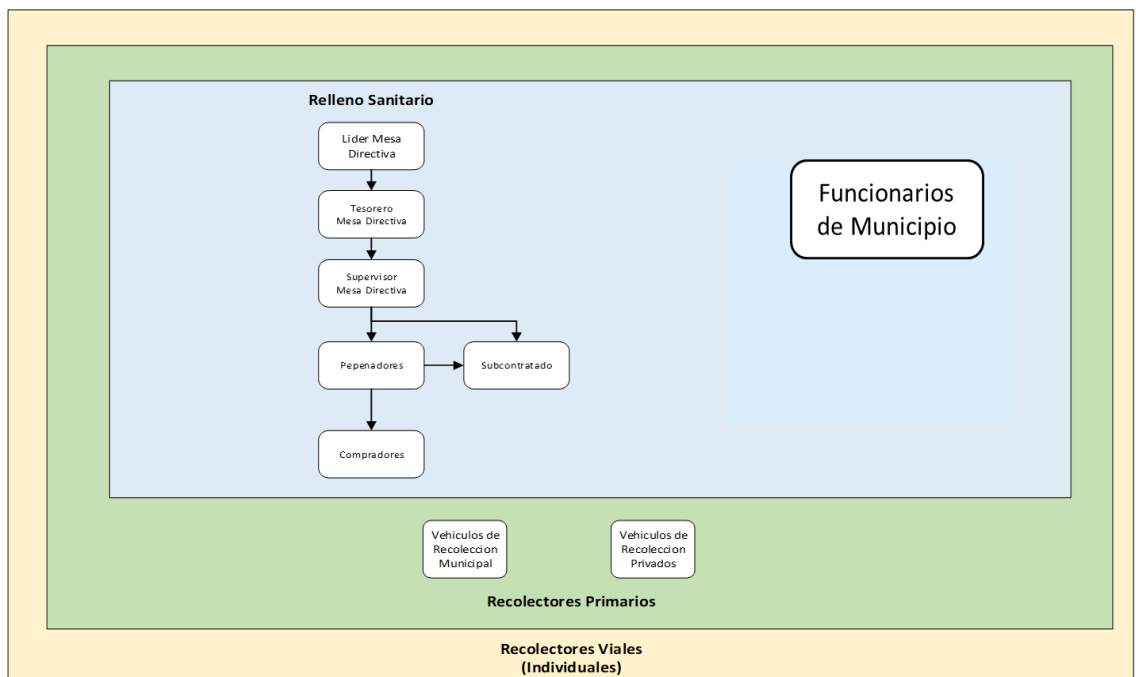

Figura 2. Mapa de actores.

El grupo de pepenadores que realizan la actividad de la pepena es de aproximadamente 380 personas, los cuales en su mayoría tienen parentesco. El estudio tiene como objetivo mostrar una visión amplia de las características sociales, de salud, educación y laboral, para su dignificación. En las siguientes figuras se podrán observar las características de situaciones específicas identificadas en el estudio de los pepenadores bajo consideraciones expuestas de algunos autores en la literatura. Esta visión permitirá comprender la heterogénea situación de sus características y el efecto de este tipo de empleo informal la ciudad de Chihuahua.

Composición del grupo de los pepenadores. El tema de género en la composición de los pepenadores es predominado por el sexomasculino con un 82% (164 personas), con un 18% restante (37 personas) representado por el sexo femenino. Los rangos de edades predominantes del grupo van de los 15 años 61 en donde la actividad pasa de forma generacional como una tradición familiar de herencia. Es decir, el 77.86% de los pepenadores tienen de 1 a 4 familiares en elmismo lugar de trabajo y el 22.13% comparten trabajo con más de cinco familiares.

La siguiente figura (3) muestra la cantidad de dependientes económicos por cada pepenador, 79 pepenadores (39.30%) se encuentran dentro del rango de 2 a 3 dependientes a su cargo y se observa que seguido a estas cantidades se encuentran34 pepenadores (16.91%) sin dependientes económicos.

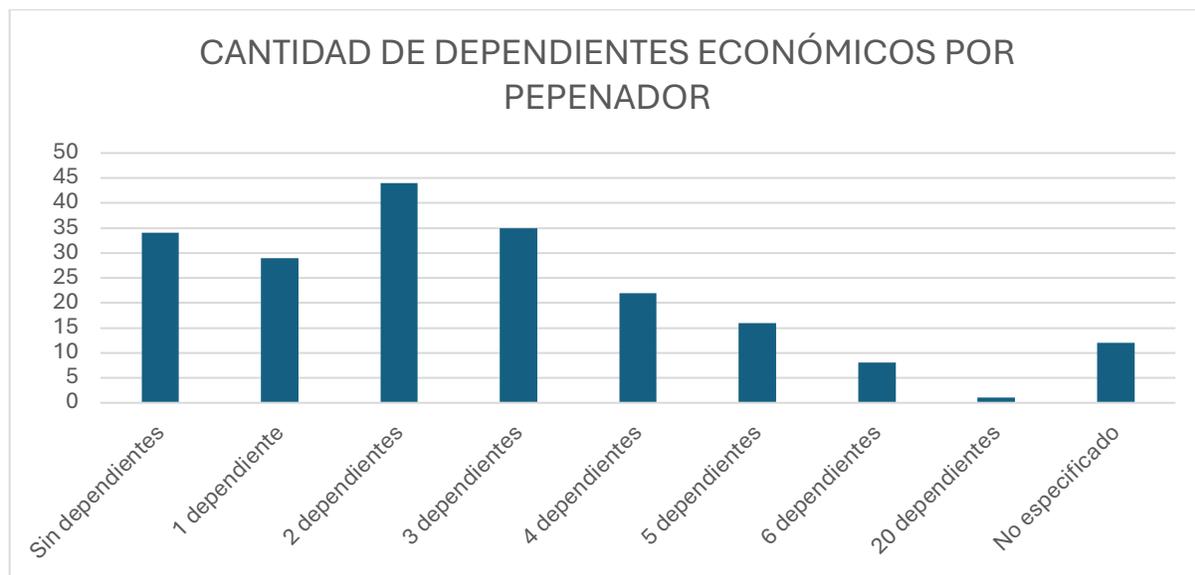

**Figura 3.** Cantidad de dependientes económicos por pepenador en la ciudad deChihuahua.

En el lugar de la disposición final considerado el espacio de trabajo de pepenadores existe una estrecha relación entre los trabajadores, ya que se organizan por familia o grupos de ellas, en materia al estudio más del 60% de los pepenadores tienen parentesco con otro trabajador, lo cual explica, la premisa anteriormente expuesta que esta manera de ganarsela vida es por tradición y lleva 35 años de antigüedad como un "negocio familiar".

Nivel educativo y formación laboral del grupo de los pepenadores

En materia de educación, el 68.15% de pepenadores concluyeron algún nivel de estudios, 86 pepenadores tuvieron oportunidad de concluir en nivel secundaria, seguido de 41 pepenadores de nivel primaria, 9 pepenadores en estudios nivel medio superior y un pepenador se graduó de técnico; el 27.36% de los pepenadores dejaron incompletos sus niveles de estudios dedicándose a la actividad de la pepena. El 93% de los pepenadores decidieron no continuar sus estudios por motivos personales, familiares y/o económicos, así mismo mencionan que la educación es inaccesible y no les permite adquirir conocimientos para su desarrollo. Por otra parte, un 39.80% de los pepenadores han adquirido capacitación en algún otro oficio y lo han ejercido, predominando la albañilería, mecánica, herrería y carpintería.

Actividad laboral en el espacio de la disposición final en Chihuahua. Derivado de la alta tasa de desempleo en la región, la forma de sobrevivencia de los pepenadores es auto emplearse en la recolección y separación de los residuos sólidos, dentro de un sector informal en condiciones de altos riesgos y con exclusión ante la legislación laboral. Ellos determinan sus jornadas laborales y días en que asisten al sitio de disposición final; en los extremos 124 pepenadores laboran 6 días a la semana, mientras que el resto de pepenadores asiste de 3 a 5 días a la semana. Logrando diariamente un intervalo de 7 a 12 horas trabajadas.

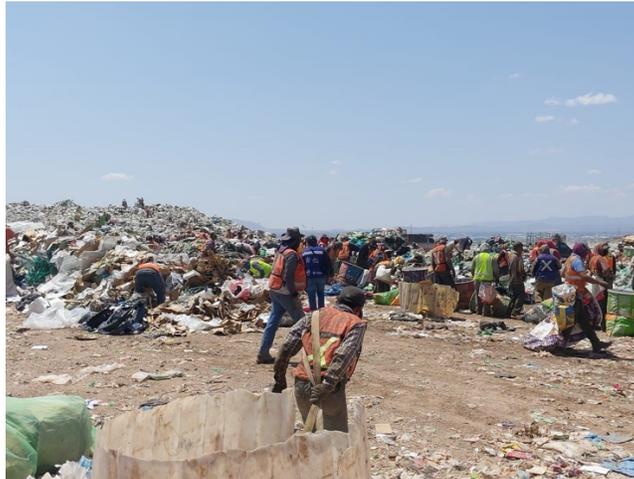

Figura 4. Desarrollo de la actividad de la pepena en la Ciudad de Chihuahua.

Recolección de los residuos sólidos urbanos. Según los datos obtenidos de los pepenadores, en cuanto a la recolección y clasificación de los residuos, el principal material recolectado es el PET con 59% de pepenadores, seguido del cartón con el 22%, 21 pepenadores deciden recolectar entre 2 y 3 residuos, como pueden ser el PET-aluminio, PET-fierro, aluminio-fierro, PET-cartón, plástico-aluminio, cartón-fierro, papel-aluminio y fierro, un 2% recolecta aluminio y en un menor porcentaje recolectan material de fierro, ropa, plástico, cobre y archivo. Se considera que la elección del material a recolectar depende de la organización, de la cantidad existente en el espacio, la demanda del mercado por producto y de la cantidad pagada por kilogramo de material.

En cuanto a la cantidad recolectada diariamente por pepenador de material, la tabla 1 indica que el 80% de los pepenadores recolectan entre 1 a 399 kilogramos diarios, el 10.94% recolectan en un rango de 400 a 799 kilogramos, el 2.48% recolectan más de 800 kilogramos y, por último, el 6.46% no especifico su cantidad.

Tabla 1. Cantidad recolectada diariamente por pepenador (kilogramos).

| CANTIDAD EN KILOGRAMO (kg) | CANTIDAD DE PEPENADORES |
|---|---|
| 1-99 kg | 21 |
| 100-199 kg | 85 |
| 200-299 kg | 40 |
| 300-399 kg | 15 |
| 400-499 kg | 8 |
| 500-599 kg | 7 |
| 600-699 kg | 3 |
| 700-799 kg | 4 |
| 800-899 kg | 2 |
| 900-999 kg | 1 |
| Más de 1,000 kg | 2 |
| No especificado | 13 |

Talleres Participativos. A través del desarrollo de talleres con aproximadamente 15 pepenadores se realizaron ejercicios de reflexión y análisis de su situación laboral actual en el relleno, en la primera dinámica que se planteó en el taller fue clasificar en "buenas" y "malas" las prácticas actuales del relleno sanitario, como resultado en malas prácticas destacan la inseguridad que sienten al trabajar junto con las máquinas y la incertidumbre por falta de comunicación con las autoridades del municipio y por el lado de buena práctica, es la libertad que sienten al ejercer este trabajo, además que el ingreso económico que reciben es diario y la experiencia en el trabajo es de mucho tiempo atrás.

En el siguiente ejercicio se realizó una lluvia de ideas sobre las condiciones que ellos aspirarían tener en su trabajo en el relleno, como lo son: continuar con su trabajo, no quieren ser empleados ni del municipio o alguna empresa, que nos les impongan un horario, no quisieran en el horario de trabajo hacer cursos de capacitación para manejo de residuos. Cabe destacar que en esta actividad reflexionaron sobre la importancia social de su actividad y la necesidad de hacer ver a la sociedad la dignidad y apoyo ambiental que ellos brindan.

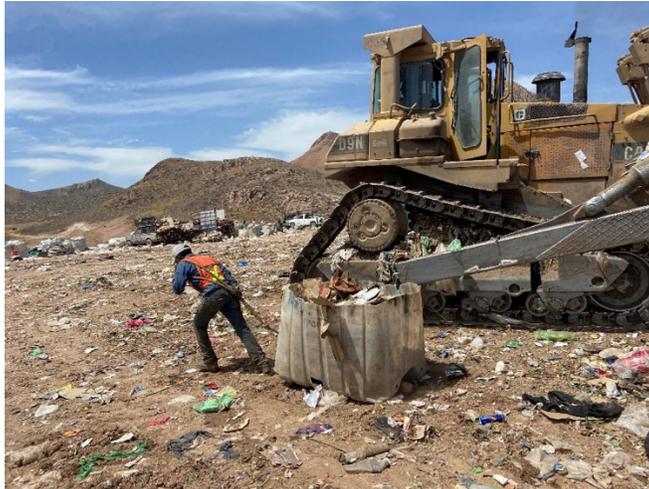

**Figura 5.** Desarrollo de la actividad de la pepena en la Ciudad de Chihuahua.

## DISCUSIÓN

De acuerdo con los datos que se obtuvieron en esta investigación, se puede confirmar lo mencionado por Santos y García (2019), quienes concluyen en su investigación que el papel de los pepenadores es fundamental en los países en vías de desarrollo. Sin embargo, siguen siendo un grupo no reconocido socialmente y que su actividad productiva sigue asociada a una explotación laboral, al rechazo social y a la vulnerabilidad.

Sin embargo, es importante reconocer el rol de los pepenadores, como lo detalla Aparicio (2021), que pone en referencia a Long (2000) y Wamsler (2000) quienes mencionan que su trabajo contribuye directamente al mejoramiento ambiental, a pesar del riesgo al que enfrentan, separan la basura que se puede reciclar y reutilizar, disminuyendo considerablemente los desechos, por lo que realizan una función que debería estar realizando la mayoría de la población.

Por tal razón, es importante reconocer y valorar la actividad de los pepenadores ya que su trabajo en condiciones precarias y de peligro, no solo implica el separar la basura, sino también en transformarla en mercancía, la cual es el medio por el cual pueden obtener un recurso económico que les permita llevar una vida digna, tanto de forma individual como de toda su familia.

Es importante mencionar que los pepenadores son un grupo que debe de ser reconocido, valorado y apoyado por las instancias gubernamentales; como lo menciona Santos y García (2019) en su estudio comprueban que en México, estos actores son invisibles e informales a comparación de los de Brasil, ya que en el país sudamericano la aplicación de acciones de estructura y organización de los pepenadores fue primordial para que avanzaran en tener una mayor visualización social y hasta tener leyes favorables que contribuyan al mejoramiento de su desempeño y contexto laboral.

Aparcana (2017) identifica hasta cinco barreras que impiden la formalización de los pepenadores: política/legal, institucional/organizacional, técnica, social, y económica/financiera, siendo las más persistentes de éstas, hasta en 75% de los casos, las que conciernen a la ausencia de políticas adecuadas y de bases legales que las sustenten.

De acuerdo con la información obtenida, se corrobora lo mencionado por De trabajo, y. C. (2018), quienes resaltan la importancia del papel económico de la recolección de residuos urbanos como una actividad laboral generadora de ingresos. Por lo que es esencial el reconocimiento social y la protección del Estado del trabajo de los pepenadores, por medio de proyectos estructurados y concretas políticas públicas y ayuden a la organización y mejoramiento de las condiciones de este grupo laboral.

## CONCLUSIÓN

La actividad de la pepena está representada en muchas partes del mundo, la población la considerada como una actividad indigna y degradante, por lo contrario, el pepenador no lo ve de esa manera, sino como una estrategia de sobrevivencia al desempleo y a la pobreza. La vulnerabilidad y la falta de reconocimiento de la pepena existe en muchos países, la ciudad de Chihuahua no es una excepción, los pepenadores son poco valorados, su labor no es reconocida y es ignorada por las autoridades, esto sin imaginar el potencial económico que genera la inclusión de este sector social. La pepena forma parte la cadena productiva del reciclaje y por lo tanto beneficia a la economía, esta actividad genera riqueza para el pepenador y cuidado para el medio ambiente, sin embargo, lamentablemente se realiza en entornos altamente contaminantes y muchas veces a bajo costo.

El grupo se compone por 380 personas, las cuales son participes del porcentaje que representa el sector informal, se observa que la pepena en la ciudad de Chihuahua es un oficio diverso, precarizado, marginado, estigmatizante y deslegitimado, a pesar de esta exclusión social (desigualdades), la actividad beneficia a la población, al entorno, al territorio, al medio ambiente, a las autoridades y por supuesto al mismo pepenador. Incorporarse a fuentes de trabajo formales ya no es una opción para el pepenador, locual explica el incremento en el porcentaje de trabajadores en la pepena, a pesar de que las condiciones de trabajo estén deterioradas cada día son más las personas beneficiadas por la actividad, en Chihuahua más de 900 personas dependen de esta actividad. El estudio nos dio a conocer el perfil del pepenador en la Ciudad de Chihuahua, se destaca que la actividad de la pepena es predominada por el sexo masculino con un 82%, el 70% de los pepenadores se encuentran entre los 21 a 50 años, el 60% de los pepenadores guardan parentesco con algún otro trabajador, el 68.15% de los pepenadores tuvieron oportunidad de concluir los diferentes niveles de estudios, el principal material de recolección es el PET y cartón con el 81% de los pepenadores, de igual manera, el 80% recolectan entre 1 a 399 kilogramos diarios. Es fundamental e importante la actividad de la pepena ya que suministra las materias primas en las industrias, son 6 empresas principales a las que se les vende el material reciclable desde hace 9 años, adquieren de los pepenadores materiales como PET, cartón, metal, plástico, lamina y aluminio, compran de 40 kg hasta 35 toneladas, pagando alrededor a $1.50 a $110 pesos, generándole al pepenador un ingreso de hasta $4,500 pesos diarios. Se demuestra que más del

60% de los pepenadores cumplen con los días y horas laborables por semana que establece la Ley Federal deTrabajo, por lo cual se puede incluir dentro de un sector formal y en políticas públicas.

Debido a esto y como propuesta para eliminar la actividad no formalizada y erradicar afectaciones por la inseguridad, abusos, falta de protección y el inacceso a los servicios de salud se recomienda que el manejo de los residuos sólidos urbanos en la Ciudad de Chihuahua, especialmente la etapa del reciclaje llevada a cabo por la actividad de la pepena, sea entendida y considerada como una responsabilidad compartida entre la ciudadanía (educación ambiental), los pepenadores (formalizar suoficio) y el gobierno (políticas públicas), ya que los beneficios generados por esta actividad mejoran las condiciones de vida de los habitantes y la calidad del medio ambiente en la ciudad de Chihuahua, con ello, es importante el acercamiento por parte de las autoridades correspondientes un acercamiento a los pepenadores para adquirir el conocimiento de las problemáticas y necesidades que ellos enfrentan diariamente, para de tal manera, llevar a cabo la creación de políticas públicas que intervengan en los problemas sociales, económicos y ambientales de los pepenadores, seguido a ello,concientizar a la ciudadanía sobre la educación y cultura ambiental, construcción de programas incluyentes y de regularización del sector informal y por último la aplicaciónde normas de cuidado y de distribución en el espacio de trabajo en el sitio de disposición final para reducir los riesgos y el entorno contaminante que este presentay que afecta al pepenador. El pepenador beneficia positivamente a todos, diariamentetransforma la basura en materia prima, transforma a la sociedad que la produce, otorgatrabajo a las empresas intermediarias, favorece al desarrollo sostenible y demuestra su relación con el territorio, en cambio, lo único que el pepenador recibe es carecer dederechos laborales, resistir diariamente a la exclusión, desigualdad y cero oportunidades, sin embargo su lucha por sostener su actividad es que gracias a ella tienen la manera de subsistir y suministrar una cantidad de ingresos a su hogar para satisfacer las necesidades básicas.

## AGRADECIMIENTOS

## REFERENCIAS